\documentclass[aps,pre,preprint,showpacs]{revtex4}
\usepackage{epsf}
\usepackage{amsmath,amssymb}

\begin{document}

\title {Chaotic scattering of a quantum particle\\
weakly coupled to a very complicated background}

\date{\today}

\author{
Valentin V. Sokolov} \affiliation{Budker Institute of Nuclear
Physics, Novosibirsk, Russia; Center for Nonlinear and
Complex Systems, Universita degli Studi dell'Insubria, Como, Italy}

\begin{abstract}

Effect of a complicated many-body environment is analyzed on the
chaotic motion of a quantum particle in a mesoscopic ballistic 
structure. The dephasing and absorption phenomena are treated on 
the same footing in the framework of a model which is free of the
ambiguities inherent to earlier models. The single-particle doorway
resonance states excited via an external channel are damped not
only because of the escape onto such channels but also due to
ulterior population of long-lived background states, the resulting 
internal damping being uniquely characterized by the spreading width. 
On the other hand, the formation of the fine structure resonances 
strongly enhances the delay time fluctuations thus broadening the 
delay time distribution.
\end{abstract}

% 05.45.Mt  Semiclassical chaos (^quantum chaos~)
% 05.60.Gg  Quantum transport
% 03.65.Nk  Scattering theory
% 24.30.-v  Resonance reactions
% 73.23.-b  Electronic transport in mesoscopic systems

\maketitle

Analog experiments with open electromagnetic microwave cavities
\cite{Stoeckmann1990,Doron1990,Alt1995,Schaefer2003} on the one
hand and extensive study of the electron transport through
ballistic microstructures \cite{Huibers1998} on the other have
drawn during last decade much attention to peculiarities of
chaotic wave interference in open billiard-like set-ups. It is
well recognized by now that the statistical approach
\cite{Verbaarschot1985,Beenakker1997,Alhassid2000} based on the
random matrix theory (RMT) provides a reliable basis for
description of the universal fluctuations characteristic of such
an interference which, in particular, manifests itself in the
single-particle resonance chaotic scattering and transport
phenomena. At the same time, experiments with ballistic quantum
dots reveal appreciable deviations from the predictions of RMT,
which persist up to low temperatures indicating some loss of the
quantum coherence. The known methods
\cite{Buettiker1986,Efetov1995} of accounting for this effect
suffer deficiencies and ambiguities
\cite{McCLer1996,Brouwer1997ii} and give different results. An
artificial prescription has been suggested in \cite{Brouwer1997ii}
to get rid of uncertainties and, simultaneously, to accord the
models.

Below we propose a model of dephasing and absorption which from
the very beginning do not suffer any ambiguity. We consider the
environment (e.g. the walls of a ballistic quantum dot) to be a
complicated many-body system with a very dense energy spectrum.
The evolution of the extended system: the coupled particle and
background, is described by means of an enlarged non-Hermitian
matrix $\mbox {\boldmath ${\cal H}$}$ of order
$N\!=\!N^{(s)}\!+\!N^{(e)}$ with two blocks ${\cal
H}^{(r)}\!=\!H^{(r)}\!-\!\frac{i}{2}A^{(r)}{A^{(r)}}^{\dag},r\!=\!s,
e$ ($N^{(e)}\gg N^{(s)}$) along the main diagonal, which are mixed
by the off-diagonal coupling $N^{(e)}\times N^{(s)}$-matrix $V$.
The upper and lower diagonal blocks represent the non-Hermitian
effective Hamiltonians of the uncoupled single-particle open
system and of the environment respectively. The Hermitian matrices
$H^{(r)}$ describe the closed counterparts (the corresponding mean
level spacings satisfy the condition $D(r\!=\!s)\gg d(r\!=\!e)$)
when the matrices $A^{(r)}$ are built of the amplitudes connecting
the internal and channel states. The single-particle and
background states have, when uncoupled, no common decay channels
so that the coupling $V$ is purely Hermitian. Since the background
states have no direct access to the observable outer channels and
attain it only due to the mixing with the internal single-particle
states, the latter play the role of doorway states.

The total $M\times M$ scattering matrix ${\bf S}(E)\!=\!{\bf
I}-i{\bf A}^{\dag} \left(E-\mbox {\boldmath ${\cal
H}$}\right)^{-1}{\bf A}$ is unitary so far as all the
$M\!=\!M^{(s)}+ M^{(e)}$ open channels, observable $(s)$ as well
as hidden $(e)$, are taken into account. The transitions between
$M^{(s)}$ observable channels are described by the submatrix
$S(E)\!=\!I-i{\cal T}(E)$ where
\begin{equation}\label{T}
{\cal T}(E)={A^{(s)}}^{\dagger}{\cal
G}^{(s)}_{\textsc{d}}(E)A^{(s)}
\end{equation}
and the matrix
\begin{equation}\label{GD}
{\cal G}^{(s)}_{\textsc{d}}(E)=\frac{1}{E-{\cal H}^{(s)}-
\Sigma^{(s)}(E)}
\end{equation}
is the upper left $N^{(s)}\times N^{(s)}$ block of the resolvent
$\mbox {\boldmath ${\cal G}$}(E)\!=\! (E-$\mbox {\boldmath ${\cal
H}$}$)^{-1}$. The subscript $D$ stands in (\ref{GD}) for
"doorway". The self-energy matrix
\begin{equation}\label{Sig}
\Sigma^{(s)}(E)=V^{\dagger}\frac{1}{E-{\cal H}^{(e)}}V\equiv
V^{\dagger}{\cal G}_0^{(e)}(E)V\,.
\end{equation}
includes all virtual transitions between the doorway states and
the environment. Generally, this matrix is not Hermitian, hence
the submatrix $S$ is not unitary. The resonance spectrum $\{{\cal
E}_{\alpha}\!=\!E_{\alpha}\!-\!\frac{i}{2}\Gamma_{\alpha}\}$, i.e.
the eigenvalue spectrum of the matrix $\mbox {\boldmath ${\cal
H}$}$, is found from the equation
\begin{equation}\label{Rs}
{\mbox {$\rm Det_{(s)}$}} \left[{\cal E}-{\cal H}^{(s)}-
\Sigma^{(s)}({\cal E})
\right]=0\,.
\end{equation}

In what follows we chiefly analyze the temporal aspects of the
scattering where the influence of the background shows up in the
most full and transparent way. In particular, a straightforward
calculation gives for the Smith delay-time submatrix
$Q=-iS^{\dag}dS/dE$ in the $M^{(s)}$-dimensional subspace of the
observable channels $Q\!=\!{\bf {b^{(s)}}}^{\dag}{\bf b^{(s)}}
\!=\!{b^{(s)}}^{\dag}b^{(s)}+{b^{(e)}}^{\dag}b^{(e)} \!=\!Q^{(s)}
+ Q^{(e)}$ where the matrix amplitudes
\begin{equation}\label{bb}
{\bf b^{(s)}}(E)=\mbox {\boldmath ${\cal G}$}(E)A^{(s)}\,; \quad
b^{(s)}(E)={\cal G}^{(s)}_{\textsc{d}}(E)A^{(s)}\,; \quad
b^{(e)}(E)={\cal G}_0^{(e)}(E)Vb^{(s)}(E)
\end{equation}
have dimensions $N\times M^{(s)}$, $N^{(s)}\times M^{(s)}$ and
$N^{(e)}\times M^{(s)}$ respectively. The two contributions
$Q^{(s,e)}$ correspond to the modified due to the interaction with
the background time delay within the dot and delay because of the
virtual transitions into the background.

We further assume the coupling matrix elements $V_{\mu m}$ to be
random Gaussian variables, $\langle V_{\mu m}\rangle$=$0$;
$\langle V_{\mu m}V^{*}_{\nu
n}\rangle$=$\delta_{\mu\nu}\delta_{mn}
\frac{1}{2}\Gamma_s\frac{1}{N^{(e)}}$ where $\Gamma_s= 2\pi\langle
|V|^2\rangle/d$ is the spreading width
\cite{BohrMot1969,Mahaux1969,Sokolov1997}. The inequality $\langle
|V|^2\rangle$$\gg$$d^2$ is implied so that the interaction, though
weak, is not weak enough for perturbation theory to be valid.
Retaining the original notations also for the averaged quantities,
we find for the Smith matrix in the main $1/N^{(e)}$ approximation
$Q$=$\Lambda\,{b^{(s)}}^{\dag}b^{(s)}$=$\Lambda Q^{(s)}$ where the
factor $\Lambda$ and the doorway Green function ${\cal
G}^{(s)}_{\textsc{d}}$ are expressed in terms of the loops
$g^{(e)}(E)$=$\frac{1}{N^{(e)}}$$\rm Tr$${\cal G}_0^{(e)}(E)$ and
$l^{(e)}(E)$=$\frac{1}{N^{(e)}}$$\rm Tr$ $\left[{{\cal
G}_0^{(e)}}^{\dag}(E){\cal G}_0^{(e)}(E)\right]$ in the Hilbert
space of the background states as
\begin{equation}\label{Lam}
\Lambda(E)=1+\frac{1}{2}\Gamma_s l^{(e)}\;(E)\,;\quad
{\cal G}^{(s)}_{\textsc{d}}(E)=\left[E-\frac{1}{2}\Gamma_s g^{(e)}(E)-
{\cal H}^{(s)}\right]^{-1}\;.
\end{equation}
The spectral equation (\ref{Rs}) reduces in the same approximation
to $N^{(s)}$ similar equations originated each from an initial
single-particle resonance.

As an instructive example, we consider first an embedded in the
background isolated single-particle resonance with the bare
complex energy ${\cal
E}_0\!=\!\varepsilon_0\!-\!\frac{i}{2}\gamma_0$, which decays onto
an only channel. As to the environment whose spectrum is very
dense and rich, its states are supposed to decay through a large
number of weak statistically equivalent channels. The
corresponding decay amplitudes $A_{\mu}^{(e)}$ are random
Gaussaian with zero means and variances $\langle
A_{\mu}^{(e)}A_{\nu}^{(e')}\rangle\!=\!\delta^{ee'}
\delta_{\mu\nu}\gamma_e/M^{(e)}$. The widths $\gamma_e$ does not
fluctuate since $M^{(e)}\gg$1. The interaction $V$ redistributes
the initial widths over exact resonances as
$\Gamma_{\alpha}\!=\!f_{\alpha}\gamma_0\!+
\!(1-f_{\alpha})\gamma_e$ \cite{Sokolov1997} with the strength
function $f_{\alpha}$$\equiv$ $f(E_{\alpha})$ obeying the
condition $\sum_{\alpha}f_{\alpha}$=1.

Peculiarities of the background energy spectrum are of no
importance in what follows. Therefore we exploit the simplest
uniform model \cite{BohrMot1969,Sokolov1997} with equidistant
levels $\epsilon_{\mu}\!=\!\varepsilon_0\!+\!\mu d\!-\!
\frac{i}{2}\gamma_e$. Then the loops as well as the strength
function can be calculated explicitly \cite{Sokolov1997}.
Depending on the magnitude of the coupling there exists two
different scenarios. In the limit
$\Gamma_s\!\gg\!\gamma_0-\gamma_e$ (the natural assumption
$\gamma_e\!\ll\!\gamma_0$ is accepted throughout the paper) of
strong interaction all the individual strengths $f_{\alpha}$ are
small, $f(E_{\alpha})\!\leq\!2d/\pi\Gamma_s$, and are distributed
around the energy $\varepsilon_0$ according to the Lorentzian with
the width $\Gamma_s$, $f_{\alpha}\!\propto\!{\cal
L}_{\Gamma_s}(E_{\alpha}\!- \!\varepsilon_0)$. Thus the original
doorway state fully dissolves in the sea of the background states.
In the opposite limit of weak coupling,
$\Gamma_s\!\ll\!\gamma_0\!- \!\gamma_e$, which is the one of our
interest, the strength
$f_0\!=\!1\!-\!\Gamma_s/(\gamma_0-\gamma_e)$ remains large when
the rest of them are small again
$f(E_{\alpha})\!\leq\!2d\,\Gamma_s/\pi(\gamma_0\!-\!\gamma_e)^2$
and distributed as ${\cal
L}_{(\gamma_0\!-\!\gamma_e)}(E_{\alpha}-\varepsilon_0)$.
Therefore, only in the weak-coupling case the doorway state
preserves individuality and can be observed through the outer
channels.

First, we briefly consider the case of a stable background
$\gamma_e\!=\!0$. The matrix $S$ is then unitary since all
resonances are excited and decay only onto the outer channels. In
the single-channel scattering considered the delay time is related
to the cross section $\sigma(E)$=$\gamma_0^2|{\cal
G}^{(s)}_{\textsc{d}}(E)|^2$ as
$Q(E)$$=$$\frac{\Lambda(E)}{\gamma_0}\sigma(E)$ and is equal to
(we set $\varepsilon_0\!=\!0$ and have taken into account that
$g^{(e)}(E)\!=\!\cot(\pi E/d)$;
$l^{(e)}(E)\!=\!(\pi/d)\sin^{-2}(\pi E/d)$)
\begin{equation}\label{Q1}
Q(E)=\gamma_0\frac{\pi\Gamma_s/2d+\sin^2\left(\pi E/d\right)}
{\left[E\sin\left(\pi E/d\right)- \frac{1}{2}\Gamma_s\cos\left(\pi
E/d\right)\right]^2+ \frac{1}{4}\gamma_0^2\sin^2\left(\pi
E/d\right)}\;.
\end{equation}
It is seen that the coupling to the background causes in both the
cross section as well as the delay time strong fine-structure
fluctuations on the scale of the background level spacing $d$. In
particular, the delay time fluctuates between
$Q(E\!=\!\epsilon_{\mu})\!=\!(2\pi/d)(\gamma_0/
\Gamma_s)\!\sim\!1/\Gamma_{\mu}$ at the points of the fine
structure levels and a much smaller value $Q(E\!\approx\!0
\!\neq\!\epsilon_{\mu})\!\approx\!(2\pi/d)(\Gamma_s/\gamma_0)$ in
between. Obviously, a particular fine-structure resonance cannot
be resolved and only quantities averaged over an energy interval
$d\!\ll\!\delta E\!\ll\!\Gamma_s$ are observed. The averaging
yields for the time delay near the doorway energy
\begin{equation}\label{avQ1}
Q=\frac{4}{\gamma_0+\Gamma_s}+\frac{2\pi}{d}\;.
\end{equation}
The first contribution comes from the doorway resonance when the
second is just the background mean level density.

Let us now take into account that the excited background states
are not stable and the background resonances strongly overlap,
$\gamma_e\gg d$, even with no coupling to the doorway states. This
smears out the fine-structure fluctuations thus reducing the loop
functions to $g^{(e)}(E)\!\Rightarrow\!-i$ and
$l^{(e)}(E)\!\Rightarrow\!2/\gamma_e$ \cite{Sokolov1997}. As
opposed to the previous case, the unitarity of the scattering
submatrix $S$ is broken. The structures narrower than $\gamma_e$
are excluded and all particles which delayed for a time larger
than $1/\gamma_e$ are irreversibly absorbed and lost from the
outgoing flow.

There exist two different temporal characteristics of the
scattering process (\ref{T}). The first one is the decay rate
$\gamma_0\!+\!\Gamma_s$ of the doorway state once excited through
the incoming channel. This rate is readily seen from the
scattering amplitude ${\cal
T}(E)\!=\!\gamma_0\!\left[E-\varepsilon_0+\frac{i}{2}\left(\gamma_0+
\Gamma_s\right)\right]^{-1}$. Since the doorway state is not an
eigenstate of the total effective Hamiltonian $\mbox {\boldmath
${\cal H}$}$ it fades out not only because the particle returns
into the outer channel but also due to the internal transitions
with formation of the exact fine structure resonances over which
the doorway state spreads. During the time $t_s\!=\!1/\Gamma_s$,
the background absorbs the particle. After this, the particle can
evade via one of the $e$ channels or be after a while emitted back
into the dot and finally escape onto the outer channel. Since the
particle reemitted by the chaotic background carries no phase
memory, the time $t_s$ is just the characteristic time of
dephasing. All the resonances, save the broad one with the width
$\Gamma_0$=$ \gamma_0-\Gamma_s$, have rather small widths and
decay much slower. Just the resonances with the widths within the
interval $\gamma_e\!<\!\Gamma_{\alpha}\!\ll\!\Gamma_s$ contributes
principally in the Wigner delay time which shows how long the
excited system still returns particles into the initial channel.
As a result, the delay time near the doorway energy equals
\begin{equation}\label{Q1s}
Q(E=\varepsilon_0)=\frac{4\gamma_0}{\left(\gamma_0+
\Gamma_s\right)^2}\Lambda
\end{equation}
where the enhance factor $\Lambda$=1+$\Gamma_s/\gamma_e$. This
factor is missing in the oversimplified consideration of
\cite{SavSomm2003} which implies that interaction with an
environment yields only absorption.

Returning to the general consideration, we as usual model the
unperturbed chaotic single-particle motion by the random matrix
theory. The observable channels are considered below to be
statistically equivalent and, to simplify the calculation, no
time-reversal symmetry is suggested. We suppose below that
$d\!\ll\!\gamma_e\!\ll\!\min \left(D,\Gamma_s\right)\!\lll\!1$
(the radius of the semicircle). If, on the contrary,
$\gamma_e\!\gg\!\Gamma_s$ the transitions into the background
become equivalent to irreversible decay similar by its properties
to the decay into continuum. Only in such a limit of full
absorption the factor $\Lambda\rightarrow 1$ and the approach of
\cite{SavSomm2003} is justified.

It is readily seen that to account for the interaction with the
background the substitution
$\varepsilon\!\Rightarrow\!\varepsilon\!-\!i\Gamma_s$ should be
done while calculating the two-point S-matrix correlation function
$C(\varepsilon)=S(E)\otimes S^{\dag}(E+\varepsilon)$. This
immediately yields the connection $C_V(t)\!=\!\exp(-\Gamma_s
t)C_0(t)$ between the Fourier transforms with and without
interaction, the additional damping being as before caused by the
spreading over the fine structure resonances. Contrary to the
models with additional fictitious channels the exponential factor
is an unambiguous consequence of our model.

The modification given in \cite{SavSomm2003} of the method
proposed in \cite{Sommers2001} allows us to calculate also the
distribution ${\cal P}(q)\!=\!(\pi
M^{(s)})^{-1}$Im$\langle$tr$(q-Q-i0)^{-1}\rangle$ proper delay
times (eigenvalues of the Smith matrix). The corresponding
generating function is proportional to the ratio of the
determinants of two $2N^{(s)}\times 2N^{(s)}$ matrices with the
following structure (compare with \cite{SavSomm2003})
\begin{eqnarray}\label{A}
A(z)&=&-i\left(E-H^{(s)}\right)\sigma_3+\frac{1}{2}\left(AA^{\dag}+
\Gamma_s\right)-\frac{\Lambda}{z}\left(1+\sqrt{1-
\frac{\Gamma_s}{\Lambda}z}\;\sigma_1\right) \nonumber \\
&\approx&-i\left(E-H^{(s)}\right)\sigma_3+\frac{1}{2}AA^{\dag}-
\left(1+\sigma_1\right)\left(\frac{\Lambda}{z}-
\frac{\Gamma_s}{2}\right)
\end{eqnarray}
where the variable $z$ spans the complex $q$-plane. The square
root sets \cite{SavSomm2003} on the positive real axes the
restriction $q\leq\Lambda/\Gamma_s\!\approx\!1/\gamma_e$. In the
approximation of the second line valid if $\Lambda\!\gg\! 1$ we
arrive to a simple relation
\begin{equation}\label{P(tau)}
{\cal P}_V(\tau)=\frac{1}{\Lambda}{\cal P}_0(\tau_V)\;,
\end{equation}
where $\tau_V\!\equiv\!\left(\tau/
\Lambda\right)\left[1-\pi\left(\Gamma_s/\Lambda D\right)
\tau\right]^{-1}\!>\!0$ and $\tau\!=\!\frac{D}{2\pi} q$. This
relation does not hold in the asymptotic region
$\tau_V$$\rightarrow$$\infty$ or $\tau$$\rightarrow$$\Lambda
D/\pi\Gamma_s$$\approx$$\frac{1}{\pi}D/\gamma_e$ where the more
elaborate rigorous expressions obtained in \cite{SavSomm2003} must
be used with the substitution $\tau$$\Rightarrow$$\tau/\Lambda$
being made. For example in the single-channel scattering with the
perfect coupling to the continuum the time delay distribution (see
\cite{FyodSomm1997}) ${\cal P}_V(\tau)\!=\!e^{-1/\tau_V}/\tau_V^3$
reaches its maximum at the point $\tau_V\!=\!1/3$ or
$\tau\!=\!(1/3)\Lambda\left(1+
\frac{\pi}{3}\Gamma_s/D\right)^{-1}\!\gg\!1/3$. The most probable
delays shift towards larger values. The estimation just made is
quantitatively valid only if $\Gamma_s\!\ll\!D/2\pi$ and the
maximum lies near the point $\frac{1}{3}\Gamma_s/\gamma_e$ distant
from the exact edge $D/2\pi\gamma_e$ of the distribution. Under
the latter restriction the approximate formula (\ref{P(tau)})
describes well the bulk of the delay time distribution which
becomes, roughly, $\Lambda$ times wider. The condition noted is
much less restrictive in the case of weak coupling to the
continuum when the transmission coefficient $T\ll 1$ and the most
probable delay time $\tau\!\approx\!T\Gamma_s/4\gamma_e\!
\ll\!D/2\pi\gamma_e$ as long as $\Gamma_s\!\ll\!D/T\pi$.

The approximation (\ref{P(tau)}) works even better when the number
of channels $M^{(s)}\!\gg\!1$ and the delay times are restricted
to a finite interval \cite{Brouwer1997,Sommers2001}. The delay
time scales in this case with ${M^{(s)}}^{-1}$ and the natural
variable looks as $\tau\!=\!qM^{(s)}D/2\pi\!=\!\Gamma_W q/T$ with
$\Gamma_W$ being the Weisskopf width. The edges of the
distribution (\ref{P(tau)}) are displace towards larger delays by
the factor $\Lambda$, $\tau_V^{(\mp)}\!=\!\Lambda\tau_0^{(\mp)}$.
One can readily convince oneself that the taken approximation
remains valid in the whole interval
$\Delta\tau_V\!=\!\Lambda(\tau_0^{(+)}\!-\!\tau_0^{(-)})$ under
condition $\Gamma_s\!\ll\!\Gamma_W$ which implies weak interaction
with the background. The width of the delay time distribution
broadens by the factor $\Lambda$ due to the influence of the
background.

At last, transport through a ballistic dot in the presence of a
background is described by off-diagonal matrix elements ${\cal
T}^{a,b}(E)$ of the matrix (\ref{T}) with the doorway Green
function $\left[E\!+\!\frac{i}{2}\Gamma_s\!-\! {\cal
H}^{(s)}\right]^{-1}$. This reproduces the results of the
imaginary-potential model \cite{Efetov1995} with
$\gamma\!=\!\Gamma_s(2\pi/D)$ as the dimensionless dephasing rate.
The unitarity is broken because of the leakage through
$e$-channels. Contrariwise, if the background states are stable no
loss of the flow takes place. This is in line with the
voltage-probe model with zero total current $I_{\phi}$
\cite{Brouwer1997ii}. It must be noted that in this case
probabilities rather than amplitudes should be averaged over the
energy to get rid of the fine-structure fluctuations.

In summary, the influence of a very complicated environment on the
chaotic single-particle scattering is analyzed. Unlike some
earlier considerations, the coupling to the background is supposed
to be purely Hermitian. Absorption takes place because of hidden
decays of the background resonance states. The single-particle
doorway states which are excited and observed through external
channels are additionally damped with the rate $\Gamma_s\!=\!
2\pi\langle |V|^2\rangle/d$ because of the spreading over the
long-lived fine structure resonances. This rate uniquely
determines the dephasing time during the particle transport
through a ballistic microstructure. The formation of the
fine-structure resonances strongly enhances delay time
fluctuations thus, in particular, broadening the distribution of
the proper delay times.

I am grateful to Y.V. Fyodorov and D.V. Savin for useful
discussions and critical remarks. The financial support by RFBR
through Grant No 03-02-16151 and through a Grant for Leading
Scientific Schools is acknowledged with thanks.

%\bibliography{../Bib/refs,../Bib/books}
%\bibliography{../Bib/refs,../Bib/books}

\begin{thebibliography}{16}
\expandafter\ifx\csname natexlab\endcsname\relax\def\natexlab#1{#1}\fi
\expandafter\ifx\csname bibnamefont\endcsname\relax
  \def\bibnamefont#1{#1}\fi
\expandafter\ifx\csname bibfnamefont\endcsname\relax
  \def\bibfnamefont#1{#1}\fi
\expandafter\ifx\csname citenamefont\endcsname\relax
  \def\citenamefont#1{#1}\fi
\expandafter\ifx\csname url\endcsname\relax
  \def\url#1{\texttt{#1}}\fi
\expandafter\ifx\csname urlprefix\endcsname\relax\def\urlprefix{URL }\fi
\providecommand{\bibinfo}[2]{#2}
\providecommand{\eprint}[2][]{\url{#2}}

\bibitem[{\citenamefont{St{\"{o}}ckmann and Stein}(1990)}]{Stoeckmann1990}
\bibinfo{author}{\bibfnamefont{H.-J.} \bibnamefont{St{\"{o}}ckmann}}
  \bibnamefont{and} \bibinfo{author}{\bibfnamefont{J.}~\bibnamefont{Stein}},
  \bibinfo{journal}{Phys. Rev. Lett.} \textbf{\bibinfo{volume}{64}},
  \bibinfo{pages}{1} (\bibinfo{year}{1990}).

\bibitem[{\citenamefont{Doron et~al.}(1990)\citenamefont{Doron, Smilansky, and
  Frenkel}}]{Doron1990}
\bibinfo{author}{\bibfnamefont{E.}~\bibnamefont{Doron}},
  \bibinfo{author}{\bibfnamefont{U.}~\bibnamefont{Smilansky}},
  \bibnamefont{and} \bibinfo{author}{\bibfnamefont{A.}~\bibnamefont{Frenkel}},
  \bibinfo{journal}{Phys. Rev. Lett.} \textbf{\bibinfo{volume}{65}},
  \bibinfo{pages}{3072} (\bibinfo{year}{1990}).

\bibitem[{\citenamefont{Alt et~al.}(1995)\citenamefont{Alt, Gr{\"a}f, Harney,
  Hofferbert, Lengeler, Richter, Schardt, and Weidenm{\"{u}}ller}}]{Alt1995}
\bibinfo{author}{\bibfnamefont{H.}~\bibnamefont{Alt}},
  \bibinfo{author}{\bibfnamefont{H.-D.} \bibnamefont{Gr{\"a}f}},
  \bibinfo{author}{\bibfnamefont{H.~L.} \bibnamefont{Harney}},
  \bibinfo{author}{\bibfnamefont{R.}~\bibnamefont{Hofferbert}},
  \bibinfo{author}{\bibfnamefont{H.}~\bibnamefont{Lengeler}},
  \bibinfo{author}{\bibfnamefont{A.}~\bibnamefont{Richter}},
  \bibinfo{author}{\bibfnamefont{P.}~\bibnamefont{Schardt}}, \bibnamefont{and}
  \bibinfo{author}{\bibfnamefont{H.~A.} \bibnamefont{Weidenm{\"{u}}ller}},
  \bibinfo{journal}{Phys. Rev. Lett.} \textbf{\bibinfo{volume}{74}},
  \bibinfo{pages}{62} (\bibinfo{year}{1995}).

\bibitem[{\citenamefont{Sch{\"{a}}fer et~al.}(2003)\citenamefont{Sch{\"{a}}fer,
  Gorin, Seligman, and St{\"{o}}ckmann}}]{Schaefer2003}
\bibinfo{author}{\bibfnamefont{R.}~\bibnamefont{Sch{\"{a}}fer}},
  \bibinfo{author}{\bibfnamefont{T.}~\bibnamefont{Gorin}},
  \bibinfo{author}{\bibfnamefont{T.}~\bibnamefont{Seligman}}, \bibnamefont{and}
  \bibinfo{author}{\bibfnamefont{H.-J.} \bibnamefont{St{\"{o}}ckmann}},
  \bibinfo{journal}{J. Phys. A: Math. Gen} \textbf{\bibinfo{volume}{36}},
  \bibinfo{pages}{3289} (\bibinfo{year}{2003}).

\bibitem[{\citenamefont{Huibers et~al.}(1998)\citenamefont{Huibers, Patel,
  Marcus, Brouwer, Duru{\"o}z, and {Harris (Jr.)}}}]{Huibers1998}
\bibinfo{author}{\bibfnamefont{A.~G.} \bibnamefont{Huibers}},
  \bibinfo{author}{\bibfnamefont{S.~R.} \bibnamefont{Patel}},
  \bibinfo{author}{\bibfnamefont{C.~M.} \bibnamefont{Marcus}},
  \bibinfo{author}{\bibfnamefont{P.~W.} \bibnamefont{Brouwer}},
  \bibinfo{author}{\bibfnamefont{C.~I.} \bibnamefont{Duru{\"o}z}},
  \bibnamefont{and} \bibinfo{author}{\bibfnamefont{J.~S.} \bibnamefont{{Harris
  (Jr.)}}}, \bibinfo{journal}{Phys. Rev. Lett.} \textbf{\bibinfo{volume}{81}},
  \bibinfo{pages}{1917} (\bibinfo{year}{1998}).

\bibitem[{\citenamefont{Verbaarschot et~al.}(1985)\citenamefont{Verbaarschot,
  Weidenm{\"{u}}ller, and Zirnbauer}}]{Verbaarschot1985}
\bibinfo{author}{\bibfnamefont{J.~J.~M.} \bibnamefont{Verbaarschot}},
  \bibinfo{author}{\bibfnamefont{H.~A.} \bibnamefont{Weidenm{\"{u}}ller}},
  \bibnamefont{and} \bibinfo{author}{\bibfnamefont{M.~R.}
  \bibnamefont{Zirnbauer}}, \bibinfo{journal}{Phys. Rep.}
  \textbf{\bibinfo{volume}{129}}, \bibinfo{pages}{367} (\bibinfo{year}{1985}).

\bibitem[{\citenamefont{Beenakker}(1997)}]{Beenakker1997}
\bibinfo{author}{\bibfnamefont{C.~W.~J.} \bibnamefont{Beenakker}},
  \bibinfo{journal}{Rev. Mod. Phys.} \textbf{\bibinfo{volume}{69}},
  \bibinfo{pages}{731} (\bibinfo{year}{1997}).

\bibitem[{\citenamefont{Alhassid}(2000)}]{Alhassid2000}
\bibinfo{author}{\bibfnamefont{Y.}~\bibnamefont{Alhassid}},
  \bibinfo{journal}{Rev. Mod. Phys.} \textbf{\bibinfo{volume}{72}},
  \bibinfo{pages}{895} (\bibinfo{year}{2000}).

\bibitem[{\citenamefont{B{\"{u}}ttiker}(1986)}]{Buettiker1986}
\bibinfo{author}{\bibfnamefont{M.}~\bibnamefont{B{\"{u}}ttiker}},
  \bibinfo{journal}{Phys. Rev. B} \textbf{\bibinfo{volume}{33}},
  \bibinfo{pages}{3020} (\bibinfo{year}{1986}).

\bibitem[{\citenamefont{Efetov}(1995)}]{Efetov1995}
\bibinfo{author}{\bibfnamefont{K.~B.} \bibnamefont{Efetov}},
  \bibinfo{journal}{Phys. Rev. Lett.} \textbf{\bibinfo{volume}{74}},
  \bibinfo{pages}{2299} (\bibinfo{year}{1995}).

  \bibitem[{\citenamefont{McCann and Lerner}(1996)}]{McCLer1996}
\bibinfo{author}{\bibfnamefont{E.}~\bibnamefont{McCann}} \bibnamefont{and}
  \bibinfo{author}{\bibfnamefont{I.~C.}~\bibnamefont{Lerner}},
  \bibinfo{journal}{J. Phys. Cond. Matter} \textbf{\bibinfo{volume}{8}},
  \bibinfo{pages}{6719} (\bibinfo{year}{1996}).

\bibitem[{\citenamefont{Brouwer and Beenakker}(1997)}]{Brouwer1997ii}
\bibinfo{author}{\bibfnamefont{P.}~\bibnamefont{Brouwer}} \bibnamefont{and}
  \bibinfo{author}{\bibfnamefont{C.}~\bibnamefont{Beenakker}},
  \bibinfo{journal}{Phys. Rev. Lett.} \textbf{\bibinfo{volume}{55}},
  \bibinfo{pages}{4695} (\bibinfo{year}{1997}).

\bibitem[{\citenamefont{Bohr and Mottelson}(1969)}]{BohrMot1969}
\bibinfo{author}{\bibfnamefont{A.}~\bibnamefont{Bohr}} \bibnamefont{and}
  \bibinfo{author}{\bibfnamefont{B.}~\bibnamefont{Mottelson}},
  \emph{\bibinfo{title}{Nuclear Structure, V.1}}
  (\bibinfo{publisher}{Benjamin}, \bibinfo{address}{New York},
  \bibinfo{year}{1969}).

  \bibitem[{\citenamefont{Mahaux and Weidenm{\"{u}}ller}(1969)}]{Mahaux1969}
\bibinfo{author}{\bibfnamefont{C.}~\bibnamefont{Mahaux}} \bibnamefont{and}
  \bibinfo{author}{\bibfnamefont{H.~A.} \bibnamefont{Weidenm{\"{u}}ller}},
  \bibinfo{book}{{\it Shell-model approach to Nuclear Reactions}}
(\bibinfo{year}{North-Holland, Amstredam, 1969}).

\bibitem[{\citenamefont{Sokolov and Zelevinsky}(1997)}]{Sokolov1997}
\bibinfo{author}{\bibfnamefont{V.~V.} \bibnamefont{Sokolov}} \bibnamefont{and}
  \bibinfo{author}{\bibfnamefont{V.}~\bibnamefont{Zelevinsky}},
  \bibinfo{journal}{Phys. Rev. C} \textbf{\bibinfo{volume}{56}},
  \bibinfo{pages}{311} (\bibinfo{year}{1997}).

\bibitem[{\citenamefont{Savin and Sommers}(2003)}]{SavSomm2003}
\bibinfo{author}{\bibfnamefont{D.~V.} \bibnamefont{Savin}}
\bibnamefont{and}
\bibinfo{author}{\bibnamefont{H.-J.} \bibnamefont{Sommers}},
\bibinfo{journal}{cond-mat/0303083}
(\bibinfo{year}{2003}).

%\bibitem[{\citenamefont{Sch{\"{a}}fer et~al.}(2003)\citenamefont{Sch{\"{a}}fer,
%  Gorin, Seligman, and St{\"{o}}ckmann}}]{Schaefer2003}
%\bibinfo{author}{\bibfnamefont{R.}~\bibnamefont{Sch{\"{a}}fer}},
%  \bibinfo{author}{\bibfnamefont{T.}~\bibnamefont{Gorin}},
%  \bibinfo{author}{\bibfnamefont{T.}~\bibnamefont{Seligman}}, \bibnamefont{and}
%  \bibinfo{author}{\bibfnamefont{H.-J.} \bibnamefont{St{\"{o}}ckmann}},
%  \bibinfo{journal}{J. Phys. A: Math. Gen} \textbf{\bibinfo{volume}{36}},
%  \bibinfo{pages}{3289} (\bibinfo{year}{2003}).

\bibitem[{\citenamefont{Sommers et~al.}(2001)\citenamefont{Sommers, Savin, and
  Sokolov}}]{Sommers2001}
\bibinfo{author}{\bibfnamefont{H.-J.} \bibnamefont{Sommers}},
  \bibinfo{author}{\bibfnamefont{D.~V.} \bibnamefont{Savin}}, \bibnamefont{and}
  \bibinfo{author}{\bibfnamefont{V.~V.} \bibnamefont{Sokolov}},
  \bibinfo{journal}{Phys. Rev. Lett.} \textbf{\bibinfo{volume}{87}},
  \bibinfo{pages}{094101} (\bibinfo{year}{2001}).

\bibitem[{\citenamefont{Brouwer et~al.}(1997)\citenamefont{Brouwer, Frahm, and
  Beenakker}}]{Brouwer1997}
\bibinfo{author}{\bibfnamefont{P.~W.} \bibnamefont{Brouwer}},
  \bibinfo{author}{\bibfnamefont{K.~M.} \bibnamefont{Frahm}}, \bibnamefont{and}
  \bibinfo{author}{\bibfnamefont{C.~W.~J.} \bibnamefont{Beenakker}},
  \bibinfo{journal}{Phys. Rev. Lett.} \textbf{\bibinfo{volume}{78}},
  \bibinfo{pages}{4737} (\bibinfo{year}{1997}).

\bibitem[{\citenamefont{Fyodorov and Sommers}(1997)}]{FyodSomm1997}
\bibinfo{author}{\bibfnamefont{Y.~V.} \bibnamefont{Fyodorov}} \bibnamefont{and}
  \bibinfo{author}{\bibfnamefont{H.-~J.}  \bibnamefont{Sommers}},
\bibinfo{journal}{J. Math. Phys.} \textbf{\bibinfo{volume}{38}},
  \bibinfo{pages}{1918} (\bibinfo{year}{1997}).


\end{thebibliography}

\end{document}